\documentclass[%
 reprint,
superscriptaddress,
 amsmath,amssymb,
 aps,
]{revtex4-2}

\usepackage{graphicx}
\usepackage{dcolumn}
\usepackage{bm}
\usepackage{multirow}
\usepackage{xcolor}
\usepackage{comment}
\usepackage{hyperref}
\newcommand{\um}{\mu \mathrm{m}}

\begin{document}

\preprint{APS/123-QED}

\title{Current-dipole detection of microscale metal defects\\}

\author{Eikichi Kimura}
\affiliation{School of Engineering, Institute of Science Tokyo, Yokohama, Kanagawa 226-8501, Japan}

\author{Eisuke Oba}
\affiliation{School of Engineering, Institute of Science Tokyo, Yokohama, Kanagawa 226-8501, Japan}

\author{Yu Saito}
\affiliation{School of Engineering, Institute of Science Tokyo, Yokohama, Kanagawa 226-8501, Japan}

\author{Yuto Yamakawa}
\affiliation{School of Engineering, Institute of Science Tokyo, Yokohama, Kanagawa 226-8501, Japan}

\author{\\El Mustapha Mansouri}
\affiliation{School of Engineering, Institute of Science Tokyo, Yokohama, Kanagawa 226-8501, Japan}

\author{Takatada Saito}
\affiliation{Research and Innovation Center, Mitsubishi Heavy Industries, Ltd., Minato, Tokyo 108-8015, Japan}

\author{Seiji Goto}
\affiliation{Research and Innovation Center, Mitsubishi Heavy Industries, Ltd., Minato, Tokyo 108-8015, Japan}

\author{Hiroshi Mashima}
\email{hiroshi.mashima.vt@mhi.com}
\affiliation{Research and Innovation Center, Mitsubishi Heavy Industries, Ltd., Minato, Tokyo 108-8015, Japan}

\author{Keigo Arai}
\email{arai.k.835f@m.isct.ac.jp}
\affiliation{School of Engineering, Institute of Science Tokyo, Yokohama, Kanagawa 226-8501, Japan}

\date{\today}

\begin{abstract}
An effective electric current dipole provides a compact description of localized current disturbances and has been employed across a broad range of physical systems and length scales. Despite its extensive use in macroscopic systems, its applicability to microscale material diagnostics remains unexplored. In this study, we demonstrate that the current-dipole representation provides an effective physical framework for the detection and characterization of microscopic defects in conductive materials. When an external current is applied to a metal, defects locally perturb the current distribution, thus generating an effective in-plane current dipole that yields a characteristic magnetic-field pattern. By measuring the magnetic-field distribution above the metal and fitting it with a dipole model, we reconstruct the position and effective strength of the dipole, thereby enabling the inference of the defect location and an effective defect-volume metric. Using wide-field magnetic imaging as a measurement platform, we observe a defect-associated magnetic signal from a 38-$\um$-long defect on the front surface of copper samples and identify a defect on the back surface using a 0.5-mm-thick copper plate. We further examine the applicability of this approach to magnetic materials, where permeability contrasts modify the field distribution. Our results establish current-dipole imaging as a general physical framework for magnetic detection of defects and highlight its potential for nondestructive inspection across a wide range of length scales.
\end{abstract}

\maketitle

\begin{figure*}[t]
  \centering
  \includegraphics[width=\linewidth]{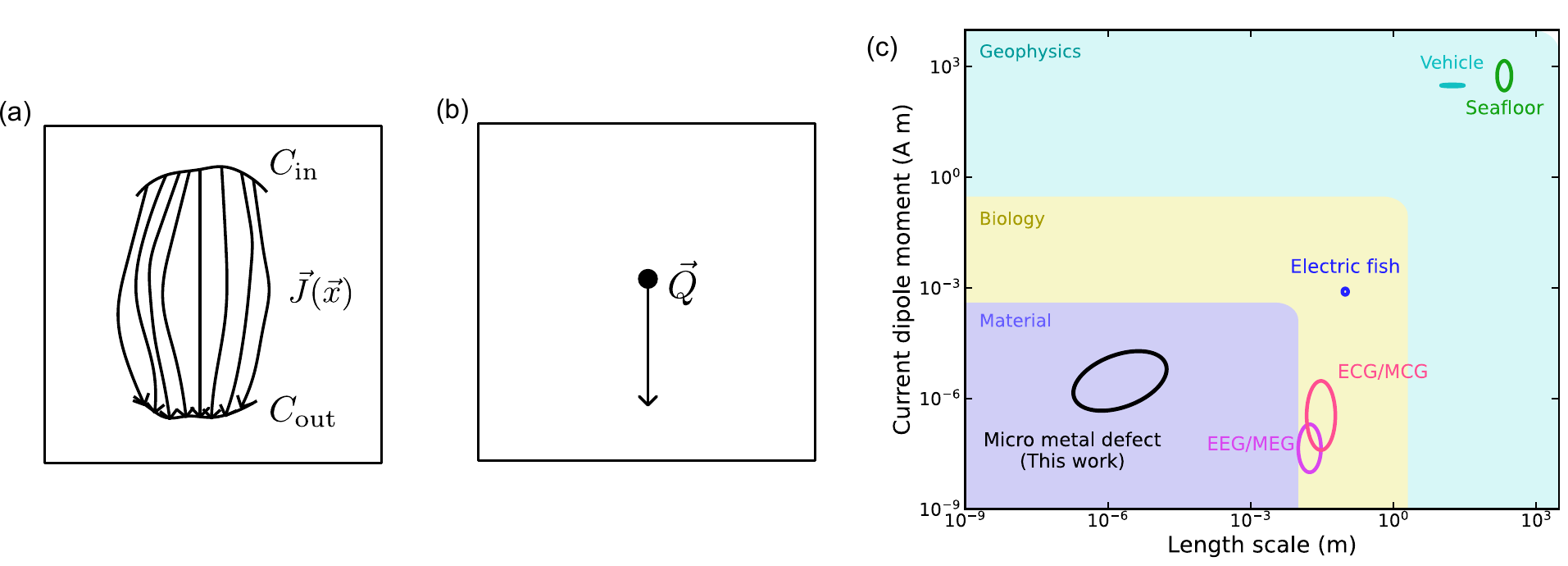}
  \caption{\label{fig:concept} Electric current dipole as a reduced representation of localized current. (a) and (b) When a localized current flows from one region ($C_{\mathrm{in}}$) to another ($C_{\mathrm{out}}$), the localized current is represented by an effective electric current dipole $\vec{Q}$. (c) Electric current dipoles appear across a wide range of systems and length scales, including seafloor hydrothermal systems~\cite{kawada_selfpotential_2018a,kawada_marine_2017}, electric signatures from vehicles and ships~\cite{woloszyn_analytical_2022}, electric fish~\cite{jun_realtime_2013a}, ECG/MCG~\cite{horigome_detection_2001a,arai_millimetrescale_2022,gonnelli_inverse_1987}, EEG/MEG~\cite{schimpf_dipole_2002a,kiebel_variational_2008,mosher_multiple_1992,uutela_global_1998}, and microscale defect-induced current perturbations, as investigated in this study.}
\end{figure*}

Across many areas of physics, localized perturbations in electric current flow can be effectively described using a current-dipole representation that relates a spatially confined perturbation in the current flow to a simple magnetic-field pattern. Although dipole models are typically introduced through far-field multipole expansions, they can also arise from a shape-based approximation in which a complex localized perturbation is replaced by an effective dipolar source that preserves the dominant magnetic-field signature relevant for inference (Figs.~\ref{fig:concept}(a) and \ref{fig:concept}(b)). This representation is used widely across diverse physical systems and length scales (Fig.~\ref{fig:concept}(c)), including seafloor hydrothermal systems and vehicle-induced electric signatures~\cite{kawada_selfpotential_2018a,kawada_marine_2017,woloszyn_analytical_2022}, electrosensory fields of electric fish~\cite{jun_realtime_2013a}, and bioelectromagnetic source modeling in electrocardiography (ECG)/magnetocardiography (MCG)~\cite{horigome_detection_2001a,arai_millimetrescale_2022,gonnelli_inverse_1987} and electroencephalography (EEG)/magnetoencephalography (MEG)~\cite{schimpf_dipole_2002a,kiebel_variational_2008,mosher_multiple_1992,uutela_global_1998}.

Despite this broad cross-scale success, current-dipole modeling has rarely been investigated as a reduced-order reconstruction framework for material defects in conductive solids at the micrometer scale. In a metal carrying an externally injected current, a localized defect causes the current to detour, creating a localized reduction in the current density, which can be modeled as an effective current dipole opposing the background flow. This representation compresses a complex current-redistribution pattern into a small set of parameters, providing a physics-informed route for inferring the defect location and an effective size metric directly from magnetic-field maps.

This capability is particularly relevant to the hydrogen energy infrastructure, where hydrogen embrittlement can initiate damage from submicrometer- to micrometer-scale defects and microcracks~\cite{birnbaum_hydrogenenhanced_1994,koyama_hydrogenassisted_2014,takai_lattice_2008a}, and monitoring their temporal evolution is critical for reliability and cost-effectiveness~\cite{kim_comprehensive_2023}. However, most inspection approaches that can resolve such small defects are destructive or challenging to deploy for repeated measurements~\cite{liu_review_2017,broberg_comparison_2015}. Thus, researchers are developing nondestructive techniques that can detect early-stage defects while enabling longitudinal monitoring.

Conventional nondestructive testing methods encounter fundamental trade-offs at these length scales. In eddy current testing, spatial resolution is limited by probe geometry and lift-off noise~\cite{garcia-martin_nondestructive_2011,long_resolution_2023}; magnetic flux leakage methods are restricted primarily to ferromagnetic materials~\cite{shi_theory_2015,feng_review_2022}; and pulse-echo ultrasonic testing can be affected by near-surface ``dead zones'' and resolution constraints imposed by pulse bandwidth and material sound speed~\cite{kiefer_simultaneous_2017}. These limitations have motivated the development of alternative modalities that can detect microscale near-surface defects with high sensitivity and minimal invasiveness.

Here, we introduce a current-dipole detection framework for microscale metal defects based on magnetic-field mapping under an externally injected current and experimentally test it using wide-field diamond magnetometry based on nitrogen-vacancy (NV) centers~\cite{degen_quantum_2017a,rondin_magnetometry_2014a}. Notably, wide-field NV magnetometry enables spatially resolved magnetic-field imaging~\cite{pham_magnetic_2011,steinert_high_2010,guo_widefield_2024,zhou_imaging_2021} and has been applied to current-density reconstruction~\cite{broadway_improved_2020,midha_optimized_2024,basso_electric_2023,chatzidrosos_eddycurrent_2019}. In this study, we (i) detect a 38-$\mu$m-long defect in copper; (ii) achieve subsurface detection by sensing defects from the opposite side of a 0.5-mm-thick copper sample; and (iii) identify a distinct permeability-dominated regime in SPCC steel, delineating the boundary of simple current-dipole inversion and facilitating extensions for ferromagnetic materials. Our results extend a widely used cross-scale physical concept to microscale defect metrology in conductive materials and suggest a route toward nondestructive inspection of metallic infrastructure.

\begin{figure}[h]
\includegraphics[width=\linewidth]{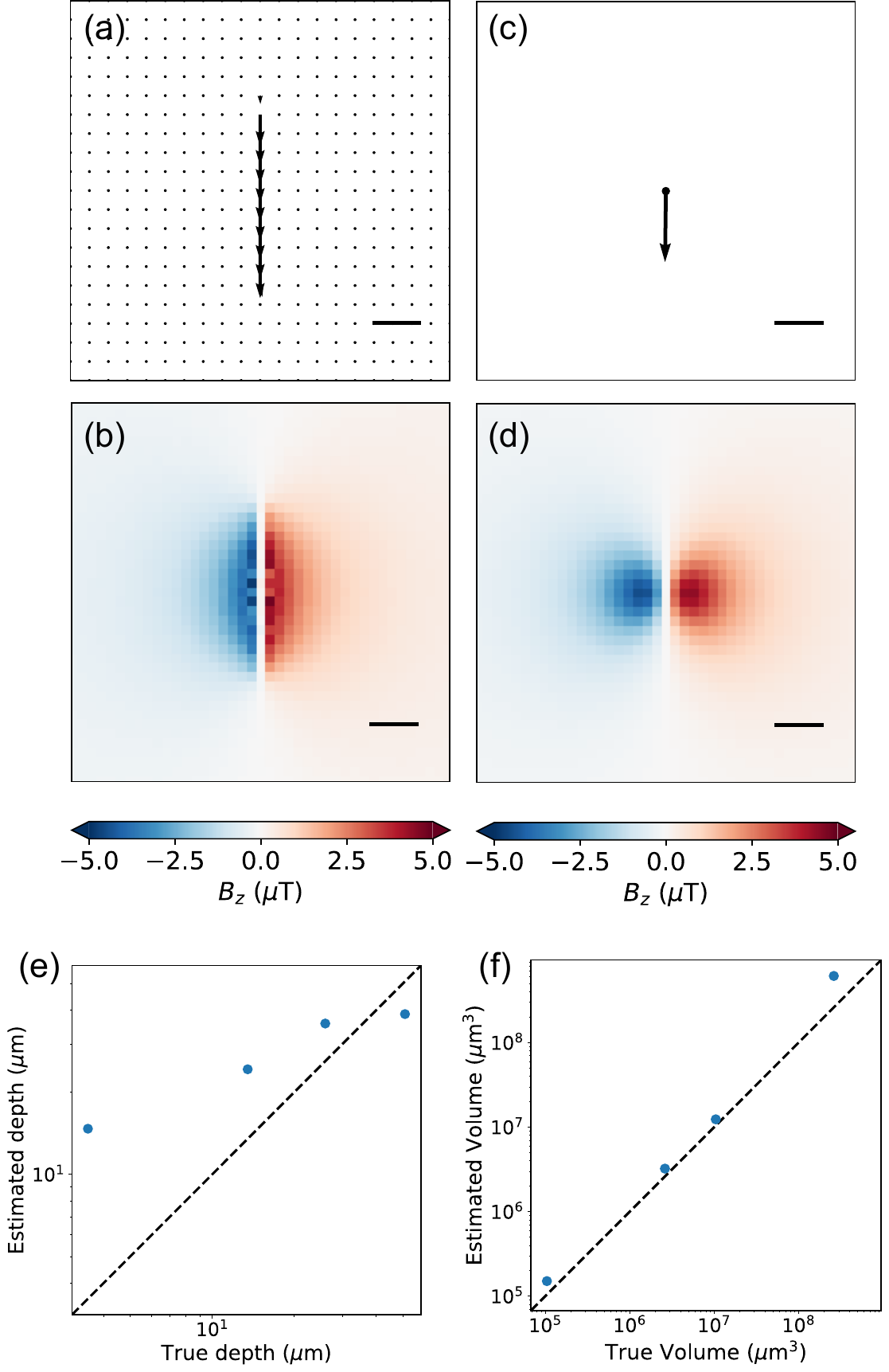}
\caption{\label{fig:workflow}
Comparison between magnetic-field distributions generated by the defect-induced current perturbation and its current-dipole representation.
(a) Current-density perturbation around the defect obtained via finite-element simulation. Black arrows indicate the current-density perturbation obtained after subtracting the mean current density within the field of view. (b) Out-of-plane magnetic field generated by the current-density perturbation in (a). (c) Effective current dipole representing the localized current perturbation in (a). (d) Out-of-plane magnetic field generated by current dipole in (c). Scale bars correspond to $50~\mu\mathrm{m}$ for distance, $1.0\times10^{-7}~\mathrm{A\,m}$ for current-dipole moment, and $1.5\times10^{6}~\mathrm{A/m^2}$ for current density.
(e) and (f) Comparison of estimated depth and effective volume with geometric reference depth and geometric volume, respectively. Dashed lines indicate the identity relation.
}
\end{figure}

We modeled a localized defect in a conducting plate as an effective current dipole embedded in a uniform background current. When a direct current is injected into a homogeneous conductor, the current density at position $\vec{x}$ in the metal can be decomposed as
\begin{eqnarray}
  \vec{J}(\vec{x}) = \vec{J}_0 + \delta\vec{J}(\vec{x}),
\end{eqnarray}
where $\vec{J}_0$ is the uniform background current density and $\delta\vec{J}(\vec{x})$ is the defect-induced perturbation. This perturbation is summarized by an effective current-dipole moment
\begin{eqnarray}
\vec{Q}(\vec{x}_0) := \int_G \delta\vec{J}(\vec{x})\,d^3x,
\end{eqnarray}
where $G$ denotes a local volume enclosing the current perturbation centered at $\vec{x}_0$. The resulting magnetic-field perturbation at position $\vec{X}$ on the observation surface is approximated by
\begin{eqnarray}
  \delta\vec{B}(\vec{X}) \simeq \frac{\mu_0}{4\pi}\vec{Q}(\vec{x}_0)\times\frac{\vec{X}-\vec{x}_0}{|\vec{X}-\vec{x}_0|^3}.
\end{eqnarray}
In our experiments, the defect suppresses core conduction and redirects the current around its perimeter. Consistent with this, the effective dipole is expected to be antiparallel to $\vec{J}_0$. For a quantitative interpretation, we parameterize $\vec{Q}$ by an effective defect volume $V_{\rm eff}$ via $\vec{Q}(\vec{x}_0)\approx -\vec{J}_0\,V_{\rm eff}$, providing a compact size metric while avoiding a full three-dimensional morphology reconstruction (see the Supplementary Material for further details).

To infer the defect parameters, we represent the localized current perturbation induced by the defect as an effective in-plane current dipole. Figure~\ref{fig:workflow}(a) shows the finite-element current-density perturbation around the defect, which was obtained by subtracting the mean current density within the field of view from the local current density, whereas Fig.~\ref{fig:workflow}(c) illustrates a reduced representation of the current dipole $\vec{Q}$. The simulated out-of-plane magnetic-field map in Fig.~\ref{fig:workflow}(b) is fitted to the dipole field shown in Fig.~\ref{fig:workflow}(d). The dipole position $\vec{x}_0=(x_0,y_0,z_0)$ is determined using bounded nonlinear least squares, whereas $\vec{Q}=(Q_x,Q_y)$ is obtained by weighted linear least squares at each trial position. The effective defect volume is estimated as follows:
\begin{eqnarray}
V_{\mathrm{eff}}=\frac{|\vec{Q}|}{|\vec{J}_0|},
\end{eqnarray}
where $|\vec{J}_0|=1\,\mathrm{A/mm^2}$. Fit uncertainties are evaluated from the 95\% confidence region of the lateral dipole position. The fitting conditions, preprocessing, and uncertainty propagation are described in the Supplementary Material.

The reconstruction method was validated via finite-element simulations of elliptical-cone defects with systematically varied depths and volumes under otherwise uniform current and material conditions. The reconstructed depth followed the nominal defect depth, with deviations determined by the defect aspect ratio, whereas the reconstructed effective volume reflected the corresponding geometric volume (Figs.~\ref{fig:workflow}(e) and \ref{fig:workflow}(f)). Details of the numerical model and simulation conditions are provided in the Supplementary Material.

\begin{table}[b]
  \caption{\label{tab:defects}
  Defects used in the experiments.}
  \begin{ruledtabular}
    \begin{tabular}{l l c r r r}
      Material & Surface & \# &
      \multicolumn{1}{c}{Length (\(\mu\)m)} &
      \multicolumn{1}{c}{Width (\(\mu\)m)} &
      \multicolumn{1}{c}{Depth (\(\mu\)m)} \\
      \hline
      \multirow{3}{*}{Copper} &
      \multirow{2}{*}{Front} &
      \hypertarget{defect1}{1} & \(3.0\times10^{2}\) & \(9\times10^{1}\) & \(6\times10^{1}\) \\
      & & \hypertarget{defect2}{2} & \(38\) & \(\sim5\) & \(\sim1\) \\
      \cline{2-6}
      & \multirow{1}{*}{Back} &
      \hypertarget{defect3}{3} & \(8.2\times10^{2}\) & \(1.2\times10^{2}\) & \(2.0\times10^{2}\) \\
      \hline
      \multirow{3}{*}[-0.6ex]{\shortstack{SPCC\\steel}} &
      \multirow{3}{*}{Front} &
      \hypertarget{defect4}{4} & \(3.2\times10^{2}\) & \(1.8\times10^{2}\) & \(5\times10^{1}\) \\
      & & \hypertarget{defect5}{5} & \(2.2\times10^{2}\) & \(2.2\times10^{2}\) & \(1.5\times10^{2}\) \\
      & & \hypertarget{defect6}{6} & \(2.6\times10^{2}\) & \(1.2\times10^{2}\) & \(1.0\times10^{2}\) \\
    \end{tabular}
  \end{ruledtabular}
\end{table}

The out-of-plane magnetic-field maps were obtained using NV-based wide-field magnetometry. The metal specimens were placed in contact with a [100]-cut diamond substrate containing approximately 4.5 ppm NV centers. Magnetic fields generated by the injected current and emanating from the metal specimens were detected through optically detected magnetic resonance (ODMR) of the NV centers. The measured resonance shifts were then converted into the out-of-plane magnetic-field component, and the spatially uniform field offset within the field of view was subtracted to isolate the defect-induced magnetic field. Further details are provided in the Supplementary Material.

\begin{figure}[t]
\includegraphics[width=\linewidth]{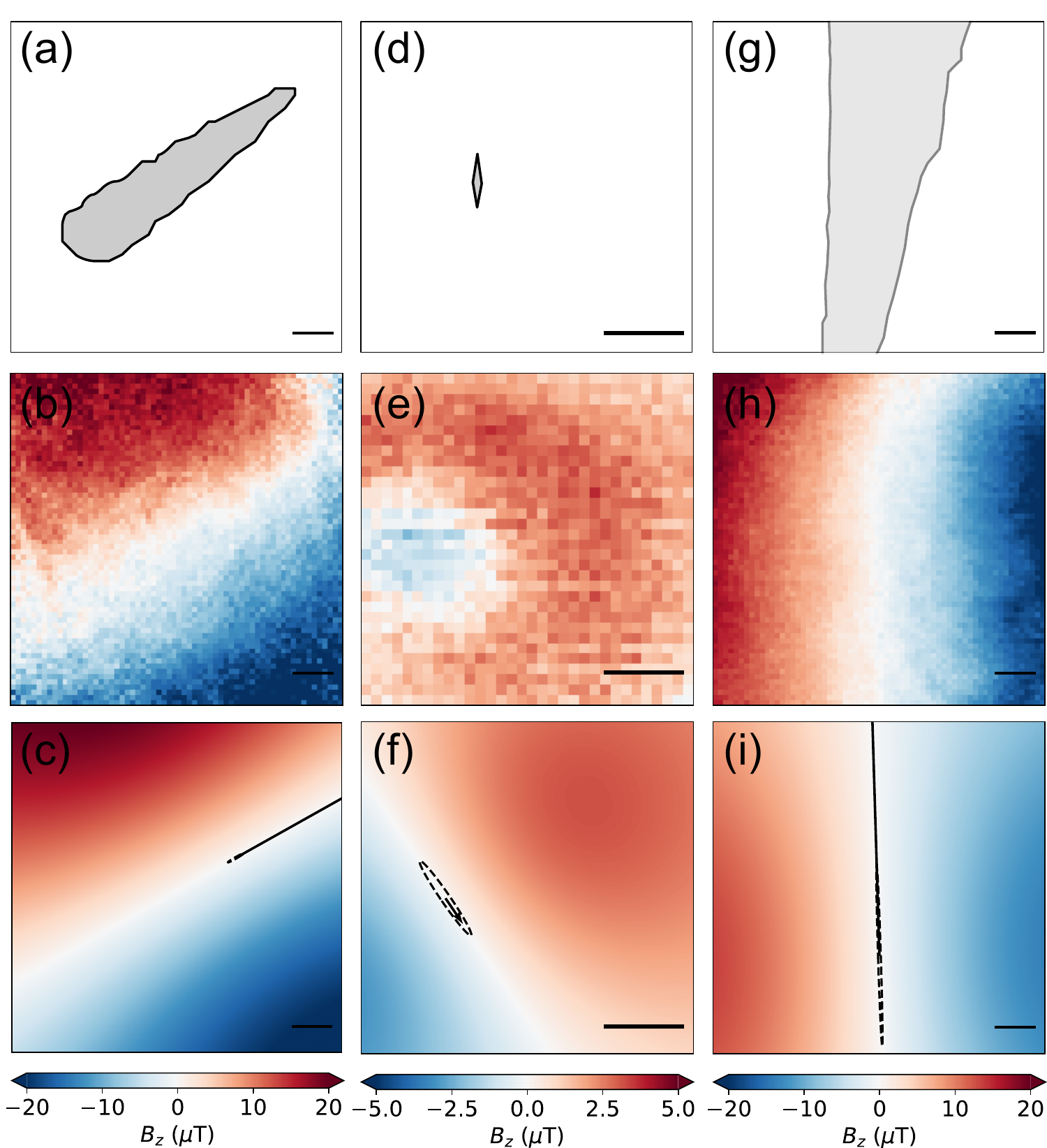}
\caption{\label{fig:Cu}
Defect reconstruction results for front-side defects (a)--(f) and a back-side defect (g)--(i) in copper specimens. Columns (a)--(c), (d)--(f), and (g)--(i) correspond to defects 1, 2, and 3, respectively. The top row shows defect outlines manually extracted from optical images, the middle row shows measured out-of-plane magnetic-field maps, and the bottom row shows the magnetic-field maps generated by fitted current dipoles. Defect dimensions are summarized in Table~\ref{tab:defects}. Scale bars correspond to $50~\mu\mathrm{m}$ in spatial distance and $5\times10^{-6}~\mathrm{A\,m}$ in current-dipole moment. In the dipole-fit maps, the origin of each arrow indicates the reconstructed lateral dipole position, while its direction and length indicate the direction and magnitude of the fitted current dipole, respectively. Dashed ellipses indicate 95\% confidence regions of reconstructed lateral dipole positions. The center of each panel was defined as the origin of the $(x,y)$ coordinates.
}
\end{figure}

We first tested the method on nonmagnetic copper, for which the measured response was dominated by the magnetic field generated by the injected current. Figures~\ref{fig:Cu}(a) and \ref{fig:Cu}(d) show the outlines of defects 1 and 2, respectively. The corresponding measured out-of-plane magnetic-field maps are shown in Figs.~\ref{fig:Cu}(b) and \ref{fig:Cu}(e), while the magnetic-field maps generated by the fitted current dipoles are shown in Figs.~\ref{fig:Cu}(c) and \ref{fig:Cu}(f). The reconstructed magnetic-field distributions closely reproduced the measured patterns. The dashed ellipses indicate the fitting uncertainty in the lateral dipole position.

Next, we evaluated subsurface defect detection using a defect introduced on the back surface of a 0.5-mm-thick copper plate. The measured out-of-plane magnetic field is broader and weaker than the fields measured for the front-side defects because of the larger source--sensor separation; nonetheless, it retains the dipolar structure characteristic of the defect-induced current redistribution (Figs.~\ref{fig:Cu}(g)--\ref{fig:Cu}(i)). The reconstructed dipole was localized near the defect position, demonstrating sensitivity to current redistribution through the 0.5-mm-thick plate. 

\begin{figure}[t]
\includegraphics[width=\linewidth]{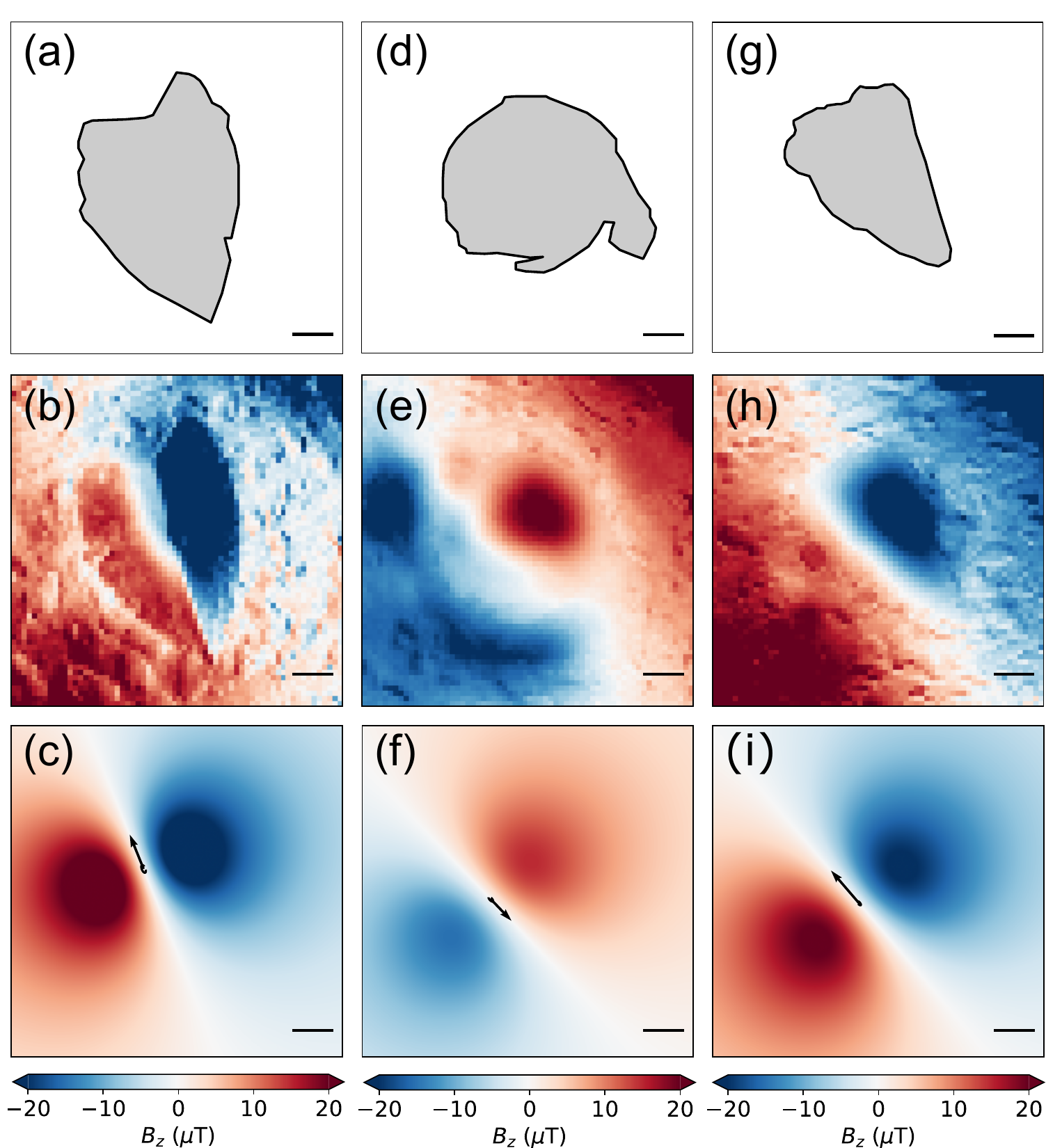}
\caption{\label{fig:Fe}
Defect reconstruction results for three front-side defects in a steel specimen. Columns (a)--(c), (d)--(f), and (g)--(i) correspond to defects 4, 5, and 6, respectively. The top row shows defect outlines manually extracted from optical images, the middle row shows measured out-of-plane magnetic-field maps, and the bottom row shows magnetic-field maps generated by fitted current dipoles. Defect dimensions are summarized in Table~\ref{tab:defects}. Scale bars correspond to $50~\mu\mathrm{m}$ in spatial distance and $5\times10^{-6}~\mathrm{A\,m}$ in current-dipole moment. In the dipole-fit maps, the origin of each arrow indicates the reconstructed lateral dipole position, while its direction and length indicate the direction and magnitude of the fitted current dipole, respectively. Dashed ellipses indicate 95\% confidence regions of reconstructed lateral dipole positions.
}
\end{figure}

Finally, we examined ferromagnetic steel, whose magnetic response is affected significantly by material permeability (Figs.~\ref{fig:Fe}(a)--(i)). The measured field patterns followed the defect geometry more closely. Interestingly, the fitted current dipoles in steel were oriented along the direction of the applied current, whereas those in copper were oriented opposite to the applied current. This difference is attributed to the ferromagnetic nature of steel, where the magnetic flux is concentrated within the high-permeability material and partially leaks around the defect~\cite{feng_review_2022}. Consequently, the simple current-dipole model did not fully reproduce the measured field distributions. Although dipole fitting remains useful for localizing defects, the reconstructed depth and effective volume cannot be directly interpreted as geometric quantities in this regime (see the Supplementary Material for the fitting conditions and results).

\begin{figure}[t!]
  \centering
  \includegraphics[width=\linewidth]{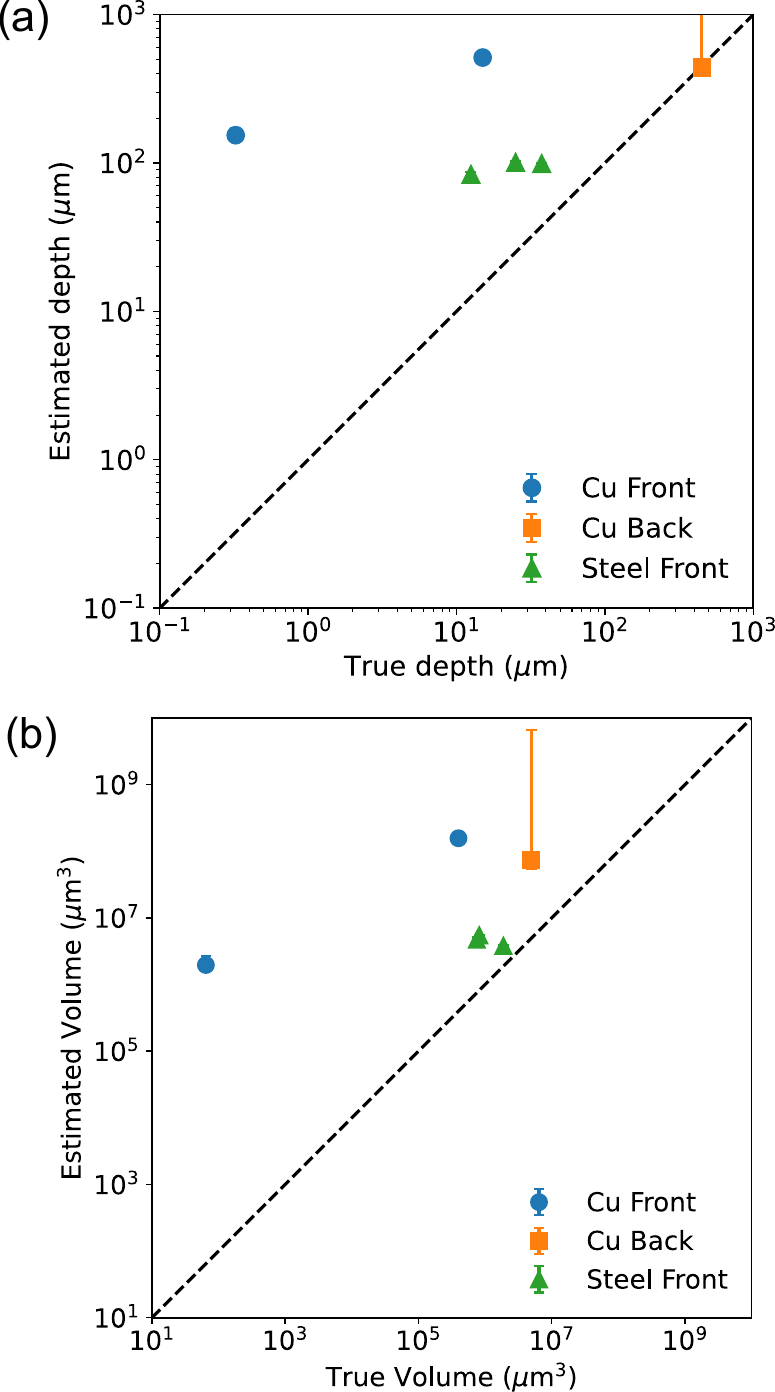}
\caption{\label{fig:size}
Comparison between reconstructed and geometric defect parameters.
(a) Reconstructed depth $z_0$ versus the geometric reference depth, defined as $h/4$ for front-surface defects and as the metal thickness minus $h/4$ for the back-side defect, where $h$ is the depth listed in Table~\ref{tab:defects}. The vertical error bars are obtained by refitting $z_0$ along the boundary of the 95\% confidence region of the lateral dipole position. For defect 3, the fit-derived depth interval is limited by the $z_0$ fitting bounds (see the Supplementary Material).
(b) Effective defect volume $V_{\mathrm{eff}}$ versus the geometric volume estimated from optical images, with vertical error bars obtained from the same boundary refits at the nominal current density.
Systematic uncertainties in the local current density and source--sensor separation are not included. Dashed lines indicate the identity relation.
}
\end{figure}

A primary limitation is the systematic uncertainty in the reconstructed depth and effective defect volume, $V_{\mathrm{eff}}=|\vec{Q}|/|\vec{J}_0|$. As shown in Fig.~\ref{fig:size}, for the front-side copper defects, the reconstructed depth deviates from the geometric reference depth by up to three orders of magnitude, whereas the effective volume differs by up to approximately four orders of magnitude. By contrast, for the back-side copper defect, the reconstructed depth is close to the geometric reference depth, while the effective volume differs by approximately one order of magnitude. Although the steel estimates appear closer to the geometric values, this apparent agreement should not be interpreted quantitatively because the simple current-dipole model does not account for the permeability-dominated magnetic response. Two systematic effects are particularly relevant for copper. The geometric reference depth is defined as the centroid depth, $h/4$, of the elliptical-cone approximation used for the front-side defects, and as the metal thickness minus $h/4$ for the back-side defect, where $h$ is the measured full defect depth listed in Table~\ref{tab:defects}. In contrast, the fitted $z_0$ represents the effective magnetic source--sensor separation; an uncharacterized gap between the sensor and specimen can therefore produce a systematic offset in depth. Likewise, $V_{\mathrm{eff}}$ is calculated using an assumed current density, $|\vec{J}_0|$, based on the bulk current density. The local current density around an individual defect may deviate from this value because of electrode geometry, finite specimen dimensions, and local current redistribution, thereby producing a systematic error in the inferred effective volume.

The vertical error bars in Fig.~\ref{fig:size} represent the fit-derived uncertainty and are distinct from these systematic effects. Specifically, the 95\% confidence region for the fitted lateral position $(x_0,y_0)$ was first determined, and $z_0$ and $|\vec{Q}|$ were then refitted along the boundary of this region. The resulting variations in $z_0$ and $V_{\mathrm{eff}}$ were used as fit-derived uncertainties. These error bars therefore quantify the precision of the inversion conditional on the adopted current-dipole model and fitting procedure, rather than the absolute accuracy of the reconstructed defect geometry.

A more fundamental limitation arises from the validity of the dipole model itself. For magnetic materials such as steel, the measured magnetic field can contain substantial contributions from permeability-dependent magnetization and flux redistribution, which are not described by the present current-dipole model. Even for nonmagnetic materials, a single dipole may not adequately represent defects with strongly asymmetric or irregular geometries, for which the associated three-dimensional current redistribution can be substantially more complex. Therefore, quantitative reconstruction of such defects requires models that account for the relevant magnetic properties and actual defect geometry.

The detectability of defects depends on the defect-induced magnetic contrast relative to the noise floor of the magnetic-field map. In our framework, detectability is governed by (i) magnetic-field sensitivity, (ii) stand-off distance between the sensor and the effective source region, and (iii) applied current density. Improving the field sensitivity can be achieved through a higher ODMR contrast, narrower linewidth, and increased fluorescence collection efficiency, which together improve the signal-to-noise ratio of the reconstructed $B_z(x_i,y_i)$ map~\cite{barry_sensitivity_2020}. In the contact configuration used here, the stand-off distance was effectively fixed by the sensor package and surface conditions. Increasing current density strengthens the magnetic signal but is limited by Joule heating in the sample and electrodes, which can degrade stability through thermal drift and may be undesirable for in situ inspection of temperature-sensitive infrastructure.

These considerations motivate frequency-selective alternating current (AC) injection as a natural next step. In DC operation, improvements primarily rely on sensor-side optimization and photon collection. In contrast, AC excitation offers two additional advantages. The skin effect concentrates the current near the metal surface, thereby increasing the near-surface current density for a given total current and enhancing the defect-induced magnetic contrast near the sensor. AC protocols also enable phase-sensitive detection that suppresses low-frequency drifts and improves the usable sensitivity; for NV-based magnetometry, AC sensing can provide substantially improved sensitivity, depending on the protocol~\cite{barry_sensitivity_2020}. Together, these effects should enable the detection of smaller defects and improve detectability.

In summary, we establish a current-dipole framework that enables nondestructive detection and parameterization of microscale metal defects from magnetic-field maps under external current injection. Although we validated the approach using wide-field NV-diamond magnetometry, the reconstruction principle is inherently sensor-agnostic and can be paired with other magnetic imaging modalities. Beyond single defects, multiple defects can be addressed by fitting superpositions of dipoles~\cite{mosher_multiple_1992}, and depth sensitivity can be gained by exploiting frequency-dependent skin depth under AC excitation~\cite{garcia-martin_nondestructive_2011}. Therefore, this framework provides a route toward nondestructive defect inspection for metallic infrastructure.

\section*{Supplementary material}
See the supplementary material for details of the experimental setup, data processing, dipole fitting, and numerical simulations.

\begin{acknowledgments}
This work was conducted as part of a collaborative research project between Mitsubishi Heavy Industries, Ltd. and the Institute of Science Tokyo and was financially supported by Mitsubishi Heavy Industries, Ltd. This work was also supported by JST PRESTO (Grant No. JPMJPR20B1). E.K. acknowledges the support from JST SPRING (Grant No. JPMJSP2180).
\end{acknowledgments}

\section*{Conflict of interests}
T.S., S.G., and H.M. are employees of Mitsubishi Heavy Industries, Ltd. This work was funded by Mitsubishi Heavy Industries, Ltd. and conducted as a collaborative research project between Mitsubishi Heavy Industries, Ltd. and the Institute of Science Tokyo. The Institute of Science Tokyo and Mitsubishi Heavy Industries, Ltd. have filed a Japanese patent application related to the technology described in this work. The authors declare no additional competing interests.

\section*{AUTHOR CONTRIBUTIONS}
E.K., E.O., and Y.S. designed the experiments, carried out the wide-field magnetic imaging, and performed data analysis. Y.Y. and E.M.M. assisted with data analysis. T.S. and S.G. prepared the metal specimens and contributed to defect fabrication. H.M. and K.A. conceived the study and supervised the project. All authors discussed the results and contributed to manuscript preparation.

\section*{Data availability}
The data that support the findings of this study are openly available in Zenodo at \url{https://doi.org/10.5281/zenodo.22198744}~\cite{zenodo}.

\bibliography{apssamp}

@article{basso_electric_2023,
  title = {Electric Current Paths in a {{Si}}:{{P}} Delta-Doped Device Imaged by Nitrogen-Vacancy Diamond Magnetic Microscopy},
  author = {Basso, Luca and Kehayias, Pauli and Henshaw, Jacob and Saleh Ziabari, Maziar and Byeon, Heejun and Lilly, Michael P. and Bussmann, Ezra and Campbell, Deanna M. and Misra, Shashank and Mounce, Andrew M.},
  year = {2023},
  journal = {Nanotechnology},
  volume = {34},
  number = {1},
  pages = {015001},
  doi = {10.1088/1361-6528/ac95a0}
}

@article{birnbaum_hydrogenenhanced_1994,
  title = {Hydrogen-Enhanced Localized Plasticity---a Mechanism for Hydrogen-Related Fracture},
  author = {Birnbaum, H. K. and Sofronis, P.},
  year = 1994,
  month = mar,
  journal = {Mater. Sci. Eng. A},
  volume = {176},
  number = {1},
  pages = {191--202},
  issn = {0921-5093},
  doi = {10.1016/0921-5093(94)90975-X}
}

@article{broberg_comparison_2015,
  title = {Comparison of {{NDT}}--Methods for Automatic Inspection of Weld Defects},
  author = {Broberg, Patrik and Sj{\"o}dahl, Mikael and Runnemalm, Anna},
  year = 2015,
  month = jan,
  journal = {Int. J. Mater. Prod. Technol.},
  volume = {50},
  number = {1},
  pages = {1--21},
  publisher = {Inderscience Publishers},
  issn = {0268-1900},
  doi = {10.1504/IJMPT.2015.066863}
}

@article{feng_review_2022,
  title = {A {{Review}} of {{Magnetic Flux Leakage Nondestructive Testing}}},
  author = {Feng, Bo and Wu, Jianbo and Tu, Hongming and Tang, Jian and Kang, Yihua},
  year = 2022,
  month = jan,
  journal = {Materials},
  volume = {15},
  number = {20},
  pages = {7362},
  publisher = {Multidisciplinary Digital Publishing Institute},
  issn = {1996-1944},
  doi = {10.3390/ma15207362},
  copyright = {http://creativecommons.org/licenses/by/3.0/},
  langid = {english}
}

@article{garcia-martin_nondestructive_2011,
  title = {Non-{{Destructive Techniques Based}} on {{Eddy Current Testing}}},
  author = {{Garc{\'i}a-Mart{\'i}n}, Javier and {G{\'o}mez-Gil}, Jaime and {V{\'a}zquez-S{\'a}nchez}, Ernesto},
  year = 2011,
  month = mar,
  journal = {Sensors},
  volume = {11},
  number = {3},
  pages = {2525--2565},
  publisher = {Molecular Diversity Preservation International},
  issn = {1424-8220},
  doi = {10.3390/s110302525},
  copyright = {http://creativecommons.org/licenses/by/3.0/},
  langid = {english}
}

@article{guo_widefield_2024,
  title = {Wide-Field {{Fourier}} Magnetic Imaging with Electron Spins in Diamond},
  author = {Guo, Zhongzhi and Huang, You and Cai, Mingcheng and Li, Chunxing and Shen, Mengze and Wang, Mengqi and Yu, Pei and Wang, Ya and Shi, Fazhan and Wang, Pengfei and Du, Jiangfeng},
  year = 2024,
  month = feb,
  journal = {Npj Quantum Inf.},
  volume = {10},
  number = {1},
  pages = {24},
  publisher = {Nature Publishing Group},
  issn = {2056-6387},
  doi = {10.1038/s41534-024-00818-9},
  copyright = {2024 The Author(s)},
  langid = {english}
}

@article{horigome_detection_2001a,
  title = {Detection of {{Cardiac Hypertrophy}} in the {{Fetus}} by {{Approximation}} of the {{Current Dipole Using Magnetocardiography}}},
  author = {Horigome, Hitoshi and Shiono, Junko and Shigemitsu, Sadahiko and Asaka, Mitsuhiro and Matsui, Akira and Kandori, Akihiko and Miyashita, Tsuyoshi and Tsukada, Keiji},
  year = 2001,
  month = aug,
  journal = {Pediatr. Res.},
  volume = {50},
  number = {2},
  pages = {242--245},
  publisher = {Nature Publishing Group},
  issn = {1530-0447},
  doi = {10.1203/00006450-200108000-00013},
  copyright = {2001 International Pediatrics Research Foundation, Inc.},
  langid = {english}
}

@article{jun_realtime_2013a,
  title = {Real-{{Time Localization}} of {{Moving Dipole Sources}} for {{Tracking Multiple Free-Swimming Weakly Electric Fish}}},
  author = {Jun, James Jaeyoon and Longtin, Andr{\'e} and Maler, Leonard},
  year = 2013,
  month = jun,
  journal = {PLOS ONE},
  volume = {8},
  number = {6},
  pages = {e66596},
  publisher = {Public Library of Science},
  issn = {1932-6203},
  doi = {10.1371/journal.pone.0066596},
  langid = {english}
}

@article{kawada_marine_2017,
  title = {Marine Self-Potential Survey for Exploring Seafloor Hydrothermal Ore Deposits},
  author = {Kawada, Yoshifumi and Kasaya, Takafumi},
  year = 2017,
  month = oct,
  journal = {Sci. Rep.},
  volume = {7},
  number = {1},
  pages = {13552},
  publisher = {Nature Publishing Group},
  issn = {2045-2322},
  doi = {10.1038/s41598-017-13920-0},
  copyright = {2017 The Author(s)},
  langid = {english}
}

@article{kawada_selfpotential_2018a,
  title = {Self-Potential Mapping Using an Autonomous Underwater Vehicle for the {{Sunrise}} Deposit, {{Izu-Ogasawara}} Arc, Southern {{Japan}}},
  author = {Kawada, Yoshifumi and Kasaya, Takafumi},
  year = 2018,
  month = aug,
  journal = {Earth Planets Space},
  volume = {70},
  number = {1},
  pages = {142},
  issn = {1880-5981},
  doi = {10.1186/s40623-018-0913-6},
  langid = {english}
}

@article{kiefer_simultaneous_2017,
  title = {Simultaneous {{Ultrasonic Measurement}} of {{Thickness}} and {{Speed}} of {{Sound}} in {{Elastic Plates Using Coded Excitation Signals}}},
  author = {Kiefer, Daniel A. and Fink, Michael and Rupitsch, Stefan J.},
  year = 2017,
  month = nov,
  journal = {IEEE Trans. Ultrason. Ferroelectr. Freq. Control},
  volume = {64},
  number = {11},
  pages = {1744--1757},
  issn = {1525-8955},
  doi = {10.1109/TUFFC.2017.2746900}
}

@article{kim_comprehensive_2023,
  title = {A {{Comprehensive Review}} on {{Material Compatibility}} and {{Safety Standards}} for {{Liquid Hydrogen Cargo}} and {{Fuel Containment Systems}} in {{Marine Applications}}},
  author = {Kim, Myung-Sung and Chun, Kang Woo},
  year = 2023,
  month = oct,
  journal = {J. Mar. Sci. Eng.},
  volume = {11},
  number = {10},
  pages = {1927},
  publisher = {Multidisciplinary Digital Publishing Institute},
  issn = {2077-1312},
  doi = {10.3390/jmse11101927},
  copyright = {http://creativecommons.org/licenses/by/3.0/},
  langid = {english}
}

@article{koyama_hydrogenassisted_2014,
  title = {Hydrogen-Assisted Decohesion and Localized Plasticity in Dual-Phase Steel},
  author = {Koyama, Motomichi and Tasan, Cemal Cem and Akiyama, Eiji and Tsuzaki, Kaneaki and Raabe, Dierk},
  year = 2014,
  month = may,
  journal = {Acta Mater.},
  volume = {70},
  pages = {174--187},
  issn = {1359-6454},
  doi = {10.1016/j.actamat.2014.01.048}
}

@article{liu_review_2017,
  title = {Review and Analysis of Three Representative Electromagnetic {{NDT}} Methods},
  author = {Liu, Shiwei and Sun, Yanhua and Gu, Min and Liu, Changde and He, Lingsong and Kang, Yihua},
  year = 2017,
  month = apr,
  journal = {Insight - Non-Destr. Test. Cond. Monit.},
  volume = {59},
  number = {4},
  pages = {176--183},
  doi = {10.1784/insi.2017.59.4.176}
}

@article{long_resolution_2023,
  title = {Resolution {{Enhanced Array ECT Probe}} for {{Small Defects Inspection}}},
  author = {Long, Cai and Zhang, Na and Tao, Xinchen and Tao, Yu and Ye, Chaofeng},
  year = 2023,
  month = jan,
  journal = {Sensors},
  volume = {23},
  number = {4},
  pages = {2070},
  publisher = {Multidisciplinary Digital Publishing Institute},
  issn = {1424-8220},
  doi = {10.3390/s23042070},
  copyright = {http://creativecommons.org/licenses/by/3.0/},
  langid = {english}
}

@article{midha_optimized_2024,
  title = {Optimized Current-Density Reconstruction from Wide-Field Quantum Diamond Magnetic Field Maps},
  author = {Midha, Siddhant and Parashar, Madhur and Bathla, Anuj and Broadway, David A. and Tetienne, Jean-Philippe and Saha, Kasturi},
  year = {2024},
  journal = {Phys. Rev. Appl.},
  volume = {22},
  number = {1},
  pages = {014015},
  doi = {10.1103/PhysRevApplied.22.014015}
}

@article{schimpf_dipole_2002a,
  title = {Dipole Models for the {{EEG}} and {{MEG}}},
  author = {Schimpf, P.H. and Ramon, C. and Haueisen, J.},
  year = 2002,
  month = may,
  journal = {IEEE Trans. Biomed. Eng.},
  volume = {49},
  number = {5},
  pages = {409--418},
  issn = {1558-2531},
  doi = {10.1109/10.995679}
}

@article{shi_theory_2015,
  title = {Theory and {{Application}} of {{Magnetic Flux Leakage Pipeline Detection}}},
  author = {Shi, Yan and Zhang, Chao and Li, Rui and Cai, Maolin and Jia, Guanwei},
  year = 2015,
  month = dec,
  journal = {Sensors},
  volume = {15},
  number = {12},
  pages = {31036--31055},
  publisher = {Multidisciplinary Digital Publishing Institute},
  issn = {1424-8220},
  doi = {10.3390/s151229845},
  copyright = {http://creativecommons.org/licenses/by/3.0/},
  langid = {english}
}

@article{takai_lattice_2008a,
  title = {Lattice Defects Dominating Hydrogen-Related Failure of Metals},
  author = {Takai, K. and Shoda, H. and Suzuki, H. and Nagumo, M.},
  year = 2008,
  month = oct,
  journal = {Acta Mater.},
  volume = {56},
  number = {18},
  pages = {5158--5167},
  issn = {1359-6454},
  doi = {10.1016/j.actamat.2008.06.031}
}

@article{woloszyn_analytical_2022,
  title = {An Analytical Four-Layer Horizontal Electric Current Dipole Model for Analysing Underwater Electric Potential in Shallow Seawater},
  author = {Woloszyn, Miroslaw and Buszman, Krystian and Rutkowski, Tomasz and Tarnawski, Jaroslaw and Rodrigo Saura, Francisco Javier},
  year = 2022,
  month = may,
  journal = {Sci. Rep.},
  volume = {12},
  number = {1},
  pages = {8727},
  publisher = {Nature Publishing Group},
  issn = {2045-2322},
  doi = {10.1038/s41598-022-12645-z},
  copyright = {2022 The Author(s)},
  langid = {english}
}

@article{degen_quantum_2017a,
  title = {Quantum Sensing},
  author = {Degen, C. L. and Reinhard, F. and Cappellaro, P.},
  year = 2017,
  month = jul,
  journal = {Rev. Mod. Phys.},
  volume = {89},
  number = {3},
  pages = {035002},
  publisher = {American Physical Society},
  doi = {10.1103/RevModPhys.89.035002}
}

@article{rondin_magnetometry_2014a,
  title = {Magnetometry with Nitrogen-Vacancy Defects in Diamond},
  author = {Rondin, L and Tetienne, J-P and Hingant, T and Roch, J-F and Maletinsky, P and Jacques, V},
  year = 2014,
  month = may,
  journal = {Rep. Prog. Phys.},
  volume = {77},
  number = {5},
  pages = {056503},
  issn = {0034-4885, 1361-6633},
  doi = {10.1088/0034-4885/77/5/056503},
  copyright = {http://iopscience.iop.org/info/page/text-and-data-mining},
  langid = {english}
}

@article{arai_millimetrescale_2022,
  title = {Millimetre-Scale Magnetocardiography of Living Rats with Thoracotomy},
  author = {Arai, Keigo and Kuwahata, Akihiro and Nishitani, Daisuke and Fujisaki, Ikuya and Matsuki, Ryoma and Nishio, Yuki and Xin, Zonghao and Cao, Xinyu and Hatano, Yuji and Onoda, Shinobu and Shinei, Chikara and Miyakawa, Masashi and Taniguchi, Takashi and Yamazaki, Masatoshi and Teraji, Tokuyuki and Ohshima, Takeshi and Hatano, Mutsuko and Sekino, Masaki and Iwasaki, Takayuki},
  year = 2022,
  month = aug,
  journal = {Commun. Phys.},
  volume = {5},
  number = {1},
  pages = {200},
  publisher = {Nature Publishing Group},
  issn = {2399-3650},
  doi = {10.1038/s42005-022-00978-0},
  copyright = {2022 The Author(s)},
  langid = {english}
}

@article{broadway_improved_2020,
  title = {Improved {{Current Density}} and {{Magnetization Reconstruction Through Vector Magnetic Field Measurements}}},
  author = {Broadway, D.A. and Lillie, S.E. and Scholten, S.C. and Rohner, D. and Dontschuk, N. and Maletinsky, P. and Tetienne, J.-P. and Hollenberg, L.C.L.},
  year = 2020,
  month = aug,
  journal = {Phys. Rev. Appl.},
  volume = {14},
  number = {2},
  pages = {024076},
  publisher = {American Physical Society},
  doi = {10.1103/PhysRevApplied.14.024076}
}

@article{pham_magnetic_2011,
  title = {Magnetic Field Imaging with Nitrogen-Vacancy Ensembles},
  author = {Pham, L M and Le Sage, D and Stanwix, P L and Yeung, T K and Glenn, D and Trifonov, A and Cappellaro, P and Hemmer, P R and Lukin, M D and Park, H and Yacoby, A and Walsworth, R L},
  year = 2011,
  month = apr,
  journal = {New J. Phys.},
  volume = {13},
  number = {4},
  pages = {045021},
  issn = {1367-2630},
  doi = {10.1088/1367-2630/13/4/045021},
  langid = {english}
}

@article{steinert_high_2010,
  title = {High Sensitivity Magnetic Imaging Using an Array of Spins in Diamond},
  author = {Steinert, S. and Dolde, F. and Neumann, P. and Aird, A. and Naydenov, B. and Balasubramanian, G. and Jelezko, F. and Wrachtrup, J.},
  year = 2010,
  month = apr,
  journal = {Rev. Sci. Instrum.},
  volume = {81},
  number = {4},
  pages = {043705},
  issn = {0034-6748},
  doi = {10.1063/1.3385689}
}

@article{barry_sensitivity_2020,
  title = {Sensitivity Optimization for {{NV-diamond}} Magnetometry},
  author = {Barry, John F. and Schloss, Jennifer M. and Bauch, Erik and Turner, Matthew J. and Hart, Connor A. and Pham, Linh M. and Walsworth, Ronald L.},
  year = 2020,
  month = mar,
  journal = {Rev. Mod. Phys.},
  volume = {92},
  number = {1},
  pages = {015004},
  publisher = {American Physical Society},
  doi = {10.1103/RevModPhys.92.015004}
}

@article{gonnelli_inverse_1987,
  title = {Inverse Problem Solution in Cardiomagnetism Using a Current Multipole Expansion of the Primary Sources},
  author = {Gonnelli, R. S. and Agnello, M.},
  year = 1987,
  month = jan,
  journal = {Phys. Med. Biol.},
  volume = {32},
  number = {1},
  pages = {133},
  issn = {0031-9155},
  doi = {10.1088/0031-9155/32/1/020},
  langid = {english}
}

@article{kiebel_variational_2008,
  title = {Variational {{Bayesian}} Inversion of the Equivalent Current Dipole Model in {{EEG}}/{{MEG}}},
  author = {Kiebel, Stefan J. and Daunizeau, Jean and Phillips, Christophe and Friston, Karl J.},
  year = 2008,
  month = jan,
  journal = {NeuroImage},
  volume = {39},
  number = {2},
  pages = {728--741},
  issn = {1053-8119},
  doi = {10.1016/j.neuroimage.2007.09.005}
}

@article{mosher_multiple_1992,
  title = {Multiple Dipole Modeling and Localization from Spatio-Temporal {{MEG}} Data},
  author = {Mosher, J.C. and Lewis, P.S. and Leahy, R.M.},
  year = 1992,
  month = jun,
  journal = {IEEE Trans. Biomed. Eng.},
  volume = {39},
  number = {6},
  pages = {541--557},
  issn = {1558-2531},
  doi = {10.1109/10.141192}
}

@article{uutela_global_1998,
  title = {Global Optimization in the Localization of Neuromagnetic Sources},
  author = {Uutela, K. and Hamalainen, M. and Salmelin, R.},
  year = 1998,
  month = jun,
  journal = {IEEE Trans. Biomed. Eng.},
  volume = {45},
  number = {6},
  pages = {716--723},
  issn = {1558-2531},
  doi = {10.1109/10.678606}
}

@misc{zenodo,
  author = {Kimura, Eikichi and Oba, Eisuke and Saito, Yu and Yamakawa, Yuto and Mansouri, El Mustapha and Saito, Takatada and Goto, Seiji and Mashima, Hiroshi and Arai, Keigo},
  title  = {Replication Data for: Current-dipole detection of microscale metal defects},
  year   = {2026},
  note   = {Zenodo, \href{https://doi.org/10.5281/zenodo.22198744}{doi:10.5281/zenodo.22198744}}
}

@article{chatzidrosos_eddycurrent_2019,
  title = {Eddy-{{Current Imaging}} with {{Nitrogen-Vacancy Centers}} in {{Diamond}}},
  author = {Chatzidrosos, Georgios and Wickenbrock, Arne and Bougas, Lykourgos and Zheng, Huijie and Tretiak, Oleg and Yang, Yu and Budker, Dmitry},
  year = 2019,
  month = jan,
  journal = {Phys. Rev. Appl.},
  volume = {11},
  number = {1},
  pages = {014060},
  publisher = {American Physical Society},
  doi = {10.1103/PhysRevApplied.11.014060}
}

@article{zhou_imaging_2021,
  title = {Imaging {{Damage}} in {{Steel Using}} a {{Diamond Magnetometer}}},
  author = {Zhou, L. Q. and Patel, R. L. and Frangeskou, A. C. and Nikitin, A. and Green, B. L. and Breeze, B. G. and Onoda, S. and Isoya, J. and Morley, G. W.},
  year = 2021,
  month = feb,
  journal = {Phys. Rev. Appl.},
  volume = {15},
  number = {2},
  pages = {024015},
  publisher = {American Physical Society},
  doi = {10.1103/PhysRevApplied.15.024015}
}


\clearpage
\onecolumngrid

\setcounter{section}{0}
\setcounter{equation}{0}
\setcounter{figure}{0}
\setcounter{table}{0}

\renewcommand{\thesection}{S\arabic{section}}
\renewcommand{\theequation}{S\arabic{equation}}
\renewcommand{\thefigure}{S\arabic{figure}}
\renewcommand{\thetable}{S\arabic{table}}

\begin{center}
{\LARGE\bfseries Supplementary Material for}\\[0.6em]
{\LARGE\bfseries Current-dipole detection of microscale metal defects}
\end{center}

\section{\label{sec:theory}Current-dipole approximation}
Two conceptually distinct current-dipole approximations are considered here: (i) a far-field approximation and (ii) a shape-based approximation. We employ the latter because the sensor--defect separation is comparable to the defect dimensions, and the defect-induced current redistribution produces a dipole-like magnetic-field pattern.
\subsection{Far-field dipole approximation}
In the absence of defects, we assume a uniform steady current density $\vec{J}_0$ in a homogeneous conductor, while the presence of a defect redistributes the current such that
\begin{eqnarray}
  \vec{J}(\vec{x}) = \vec{J}_0 + \delta\vec{J}(\vec{x}),
\end{eqnarray}
where $\delta\vec{J}(\vec{x})$ is the defect-induced current-density perturbation. By linearity of the Biot--Savart law, the magnetic field at an observation point $\vec{X}$ decomposes as
\begin{eqnarray}
  \vec{B}(\vec{X}) &=& \vec{B}_0(\vec{X}) + \delta\vec{B}(\vec{X}).
\end{eqnarray}
Specifically,
\begin{eqnarray}
  \label{eq:dB}
  \delta\vec{B}(\vec{X}) &=& \frac{\mu_0}{4\pi}
  \int_G \delta\vec{J}(\vec{x})\times\frac{\vec{X}-\vec{x}}{|\vec{X}-\vec{x}|^3}d^3x,
\end{eqnarray}
where $G$ is the volume enclosing the perturbation. By introducing defect-centered coordinates $\vec{R} := \vec{X}-\vec{x}_0$ and $\vec{\rho} := \vec{x}-\vec{x}_0$, Eq.~\eqref{eq:dB} becomes
\begin{eqnarray}
  \label{eq:rho}
  \delta\vec{B}(\vec{X})
  &=& \frac{\mu_0}{4\pi}\int_G \delta\vec{J}(\vec{x}_0+\vec{\rho})\times
  \frac{\vec{R}-\vec{\rho}}{|\vec{R}-\vec{\rho}|^3}\,d^3\rho.
\end{eqnarray}
When the spatial extent of the perturbation is much smaller than the source--sensor separation, i.e., $|\vec{\rho}|\ll |\vec{R}|$, the kernel in Eq.~\eqref{eq:rho} can be expanded about $\vec{\rho}=\vec{0}$:
\begin{eqnarray}
  \frac{\vec{R}-\vec{\rho}}{|\vec{R}-\vec{\rho}|^3}
  &=& \frac{\vec{R}}{|\vec{R}|^3} + O\left(\frac{|\vec{\rho}|}{|\vec{R}|^3}\right).
\end{eqnarray}
At zeroth order, Eq.~\eqref{eq:rho} becomes
\begin{eqnarray}
  \delta\vec{B}(\vec{X})
  &\simeq& \frac{\mu_0}{4\pi}\int_G \delta\vec{J}(\vec{x})\times
  \frac{\vec{R}}{|\vec{R}|^3}\,d^3x \\
  &=& \frac{\mu_0}{4\pi}\left(\int_G \delta\vec{J}(\vec{x})\,d^3x\right)\times
  \frac{\vec{X}-\vec{x}_0}{|\vec{X}-\vec{x}_0|^3}.
\end{eqnarray}
We then define the current dipole as
\begin{eqnarray}
  \vec{Q}(\vec{x}_0) := \int_G \delta\vec{J}(\vec{x})\,d^3x,
\end{eqnarray}
so that the defect-induced magnetic field in the far field is written as
\begin{eqnarray}
  \delta\vec{B}(\vec{X}) \simeq \frac{\mu_0}{4\pi}\vec{Q}(\vec{x}_0)\times\frac{\vec{X}-\vec{x}_0}{|\vec{X}-\vec{x}_0|^3}.\label{eq:dip}
\end{eqnarray}
Within the defect volume $D\subset G$, $\vec{J}(\vec{x})=0$, such that $\delta\vec{J}(\vec{x})=-\vec{J}_0$. Therefore, the integral over $G$ can be decomposed as
\begin{eqnarray}
  \vec{Q}(\vec{x}_0)
  &=& \int_D \delta\vec{J}(\vec{x})\,d^3x + \int_{G\setminus D} \delta\vec{J}(\vec{x})\,d^3x \nonumber\\
  &\simeq& -\vec{J}_0\,V_{D} + \int_{G\setminus D} \delta\vec{J}(\vec{x})\,d^3x,
\end{eqnarray}
where $V_D$ is the volume of $D$. Motivated by the dominant missing-current contribution, we define an effective defect volume $V_{\rm eff}$ through
\begin{eqnarray}
  \vec{Q}(\vec{x}_0)\approx -\vec{J}_0\,V_{\rm eff}.
\end{eqnarray}
For a nonmagnetic conductor in this approximation, $\vec{Q}(\vec{x}_0)$ is antiparallel to $\vec{J}_0$.

\subsection{Shape-based dipole approximation}
When the localized defect-induced current perturbation is represented by an effective point source at $\vec{x}_0$, we write
\begin{eqnarray}
  \vec{J}(\vec{x}) = \vec{J}_0 + \vec{Q}\,\delta^{(3)}(\vec{x}-\vec{x}_0),
\end{eqnarray}
where $\delta^{(3)}$ is the three-dimensional Dirac delta function. The corresponding magnetic-field perturbation is
\begin{eqnarray}
  \delta \vec{B}(\vec{X})
  &=& \frac{\mu_0}{4\pi}\int_G \vec{Q}\,\delta^{(3)}(\vec{x}-\vec{x}_0)
  \times\frac{\vec{X}-\vec{x}}{|\vec{X}-\vec{x}|^3}\,d^3x \\
  &=& \frac{\mu_0}{4\pi}\vec{Q}\times\frac{\vec{X}-\vec{x}_0}{|\vec{X}-\vec{x}_0|^3},
\end{eqnarray}
which has the same form as Eq.~\eqref{eq:dip}. This representation is used as an effective model of localized current redistribution rather than as a far-field expansion.

\section{\label{sec:numerical_modeling}Numerical modeling of the current-dipole representation}
Finite-element simulations were performed using COMSOL Multiphysics\textsuperscript{\textregistered}. The metal plate was placed within an air domain. A fixed electric potential was applied to one end face of the metal plate, while the opposite end face was grounded. At the outer boundary of the surrounding air domain, electrical insulation and magnetic insulation boundary conditions were imposed. No separate boundary condition was assigned at the metal–air interface; the interface was treated as an internal material boundary with the standard continuity conditions for the electromagnetic fields.

Magnetic-field distributions were evaluated on representative observation planes and cross sections near the defect. The out-of-plane magnetic-field maps obtained from the finite-element simulations were fitted with the same in-plane current dipole model used for the experimental data. A single initial position was used, with $x_0$ and $y_0$ set to the center of the exported field map and $z_0=100~\mu\mathrm{m}$. The only constraint imposed on the fit was the physical condition $z_0\geq0$, while $x_0$ and $y_0$ were left unconstrained.

The fitted dipole magnitude was calculated as
\begin{eqnarray}
|\vec{Q}|=\sqrt{Q_x^2+Q_y^2},
\end{eqnarray}
and converted to an effective defect volume according to
\begin{eqnarray}
V_{\rm eff}=\frac{|\vec{Q}|}{|\vec{J}_0|},
\end{eqnarray}
where the background current density was fixed at $|\vec{J}_0|=1~\mathrm{A/mm^2}$.

For comparison with the fitted depth, the reference depth of each elliptical-cone defect was defined from its centroid. For front-side defects, the centroid depth is $h/4$, where $h$ is the measured defect depth. For the back-side defect, the reference depth is the metal thickness minus $h/4$.

\section{\label{sec:setup}Experimental setup and samples}
\subsection{Wide-field NV magnetometry}
The negatively charged nitrogen-vacancy (NV) center in diamond has a spin-1 ground state whose resonance frequencies shift with the magnetic field. We measured these shifts using continuous-wave optically detected magnetic resonance (CW-ODMR) and converted the current-induced resonance shifts into magnetic-field maps.

An NV diamond plate (Element Six, DNV-B14) was placed in direct contact with the metal specimen. A 532-nm laser (Coherent, Verdi-G2) was switched using an acousto-optic modulator (Gooch \& Housego, AOMO 3080-125) and expanded using a beam expander (Thorlabs, GBE05-A). A plano-convex lens ($f=125$ mm; Thorlabs, LA1986-A) placed before a 20$\times$ objective lens (Mitutoyo, M Plan Apo SL 20$\times$) produced an excitation spot approximately 500~$\mu$m in diameter at the sample. NV fluorescence was separated from the excitation light by a dichroic mirror (Thorlabs, CM1-DCH/M), passed through a 650-nm long-pass filter (Thorlabs, FELH0650), and recorded with a CCD camera (Andor, iXon Ultra 897). The field of view was $420\times420~\mu\mathrm{m}^2$.

Microwaves were delivered by a single-loop copper-wire coil. The microwave signal was generated by a signal generator (Keysight, N5182B) gated by a switch (Mini-Circuits, ZASWA-2-50DRA+), with timing provided by a digital timing generator (Tektronix, DTG5274). A SmCo permanent magnet (14~mm outer diameter $\times$ 9~mm inner diameter $\times$ 5~mm thickness; Magfine, SR0001) positioned approximately 3 cm above the plate supplied a bias field of 3--4 mT to separate the ODMR resonances.

For each specimen, ODMR spectra were recorded with the applied DC current on and off. The nominal current density was $1~\mathrm{A/mm^2}$. Current-induced resonance shifts were converted to $B_z$ as described in Sec.~\ref{sec:dataanalysis}.

\subsection{Metal specimens and defect preparation}
The specimens consisted of 0.5-mm-thick copper and SPCC steel plates. The copper plates measured $50\times75$ mm$^2$ and $10\times50$ mm$^2$; the steel plate measured $50\times100$ mm$^2$. Copper defects were made with either a metal utility knife (MonotaRO, SX70-2N) or a pencil-type scriber needle, and steel defects were made with a diamond-wheel glass cutter (TRUSCO, TGCD-1). The defect dimensions were measured from the optical microscope images. Each plate was connected to a DC supply through a series resistor, and the applied voltage was adjusted to obtain the nominal current density stated above.

The geometric dimensions of the defects used in the experiments are listed in Table~\ref{tab:Sdefects}.
For the estimation of the geometric defect volume, each defect was approximated as an elliptical cone with length $L$, width $W$, and depth $h$. The geometric volume was calculated as
\begin{equation}
V_{\mathrm{geo}}
=
\frac{1}{3} \times
\pi \times \frac{L}{2}\times\frac{W}{2}\times h
=
\frac{\pi LWh}{12}.
\end{equation}
Here, $h$ represents the measured defect depth.
The resulting $V_{\mathrm{geo}}$ values were used as geometric reference volumes for comparison with the reconstructed effective defect volume $V_{\mathrm{eff}}$.

\begin{table*}[t]
\caption{\label{tab:Sdefects}
Geometric dimensions and estimated volumes of the defects used in the experiments.
The geometric volume $V_{\mathrm{geo}}$ was calculated from the measured dimensions by approximating each defect as an elliptical cone: $V_{\mathrm{geo}}=\pi LWh/12$. Dimensions are rounded for presentation, whereas $V_\mathrm{geo}$ was calculated using the underlying measured values.
}
\begin{ruledtabular}
\begin{tabular}{llcccc}
Defect & Dataset & Length & Width & Depth & $V_{\mathrm{geo}}$ \\
 &  & ($\mu$m) & ($\mu$m) & ($\mu$m) & ($\mu$m$^3$) \\
\hline
1 & Cu\_Front\_2 & $3.0\times10^{2}$ & $9\times10^{1}$ & $6\times10^{1}$ & $4.0\times10^{5}$ \\
2 & Cu\_Front\_1 & $38$ & $\sim5$ & $\sim1$ & $65$ \\
3 & Cu\_Back\_12 & $8.2\times10^{2}$ & $1.2\times10^{2}$ & $2.0\times10^{2}$ & $5.0\times10^{6}$ \\
4 & Fe\_Front\_1 & $3.2\times10^{2}$ & $1.8\times10^{2}$ & $5\times10^{1}$ & $7.5\times10^{5}$ \\
5 & Fe\_Front\_2 & $2.2\times10^{2}$ & $2.2\times10^{2}$ & $1.5\times10^{2}$ & $1.9\times10^{6}$ \\
6 & Fe\_Front\_3 & $2.6\times10^{2}$ & $1.2\times10^{2}$ & $1.0\times10^{2}$ & $8.2\times10^{5}$ \\
\end{tabular}
\end{ruledtabular}
\end{table*}

\section{\label{sec:dataanalysis}Data analysis and dipole fitting}

\subsection{ODMR preprocessing and magnetic-field map}
The raw image cube was spatially averaged in non-overlapping $8\times8$-pixel blocks. At each binned pixel, the spectrum was fitted to a Lorentzian function with multiple peaks,
\begin{eqnarray}
S(f)=I_0-\sum_k\frac{C_k\Gamma_k^2}{(f-f_{0,k})^2+\Gamma_k^2},
\end{eqnarray}
where $I_0$ is the fluorescence baseline and $C_k$, $f_{0,k}$, and $\Gamma_k$ are the contrast, center frequency, and half-width of the $k$th resonance, respectively. The resonance-frequency seeds used for each dataset are listed in Table~\ref{tab:odmrseeds}.

For each pixel, the lowest fitted resonance was selected from the current-on and current-off spectra. The current-induced magnetic-field signal was calculated as
\begin{eqnarray}
B^{\rm raw}(x_i,y_i)
&=&-\frac{f_{0,{\rm on}}(x_i,y_i)-f_{0,{\rm off}}(x_i,y_i)}{\gamma},\\
\gamma
&=&28\times10^9\ \mathrm{Hz/T}.
\end{eqnarray}
The spatial mean was then subtracted, and the resulting field was converted to the out-of-plane component according to
\begin{eqnarray}
B_z(x_i,y_i)
&=&
\sqrt{3}\left[B^{\rm raw}(x_i,y_i)-\left\langle B^{\rm raw}\right\rangle\right].
\end{eqnarray}
This offset subtraction removes the spatially uniform magnetic-field contribution. For a [100]-cut diamond, the projection of each NV axis onto the surface normal is $1/\sqrt{3}$ in magnitude, giving the factor of $\sqrt{3}$.

A planar background was removed only for copper front-side defect~\hyperlink{defect2}{2} because the defect-induced magnetic field was relatively small compared with the systematic background slope. For this dataset, the coefficients $(a,b,c)$ were obtained by least squares over all finite pixels of the field map,
\begin{eqnarray}
B_{\rm plane}(x,y)=ax+by+c,
\end{eqnarray}
and the dipole fit used $B_z'=B_z-B_{\rm plane}$. For all other datasets, $B_z'=B_z$. The preprocessing conditions are listed in Table~\ref{tab:fitconditions}.

\subsection{Weighted current-dipole fit}
Within the sample-specific fitting region of interest (ROI), the out-of-plane field of an in-plane current dipole is
\begin{eqnarray}
B_z^{\rm dip}(x,y)
=\frac{\mu_0}{4\pi}
\frac{Q_x(y-y_0)-Q_y(x-x_0)}{\left[(x-x_0)^2+(y-y_0)^2+z_0^2\right]^{3/2}}.
\end{eqnarray}
For a fixed position $\mathbf{r}_0=(x_0,y_0,z_0)$, this expression is linear in $\mathbf{Q}=(Q_x,Q_y)^T$. Defining a design matrix $A$ from the two geometric kernels and $W=\mathrm{diag}(\sigma_{B,i}^{-2})$, the moment is obtained from weighted linear least squares,
\begin{eqnarray}
\widehat{\mathbf{Q}}(\mathbf{r}_0)
=\left(A^TWA\right)^{+}A^TW\mathbf{b},
\end{eqnarray}
where the superscript $+$ denotes the Moore--Penrose pseudoinverse and $\mathbf{b}$ contains the measured $B_z'$ values. The position is then found by minimizing the profiled weighted residual,
\begin{eqnarray}
\chi^2(\mathbf{r}_0)=\sum_{i=1}^{N}
\left[
\frac{B'_{z,i}-B^{\rm dip}_{z,i}(\mathbf{r}_0,\widehat{\mathbf{Q}})}{\sigma_{B,i}}
\right]^2,
\end{eqnarray}
subject to the bounds in Table~\ref{tab:fitconditions}. The reduced chi-squared statistic is defined as
\begin{eqnarray}
\chi^2_\nu=\frac{\chi^2}{N-5},
\end{eqnarray}
where the five fitted parameters are $(x_0,y_0,z_0,Q_x,Q_y)$.

To reduce sensitivity to the starting point, each dataset was fitted using a $5\times5\times5$ Cartesian grid of initial $(x_0,y_0,z_0)$ values. The 125 starting points were uniformly spaced over the specified parameter bounds. Each start was optimized by bounded nonlinear least squares, and the solution with the smallest $\chi^2_\nu$ was retained. The fitting regions, parameter bounds, and preprocessing conditions are listed in Table~\ref{tab:fitconditions}.

\begin{table*}[t]
\caption{\label{tab:fitconditions}
Dipole-fitting conditions. Coordinates are relative to the center of the magnetic-field image.
}
\begin{ruledtabular}
\begin{tabular}{llcccccc}
Defect & Dataset & Fit ROI $x$ & Fit ROI $y$ & $x_0$ bounds & $y_0$ bounds & $z_0$ bounds & Plane removal \\
 &  & ($\mu$m) & ($\mu$m) & ($\mu$m) & ($\mu$m) & ($\mu$m) &  \\
\hline
1 & Cu\_Front\_2 & $[-210,210]$ & $[-210,210]$ & $[-210,210]$ & $[-210,210]$ & $[0,800]$ & No \\
2 & Cu\_Front\_1 & $[-40,50]$ & $[-80,50]$ & $[-20,30]$ & $[-60,10]$ & $[0,200]$ & Yes \\
3 & Cu\_Back\_12 & $[-70,70]$ & $[-70,70]$ & $[-100,100]$ & $[-100,100]$ & $[400,2000]$ & No \\
4 & Fe\_Front\_1 & $[-105,105]$ & $[-105,105]$ & $[-105,105]$ & $[-105,105]$ & $[0,500]$ & No \\
5 & Fe\_Front\_2 & $[-105,105]$ & $[-105,105]$ & $[-105,105]$ & $[-105,105]$ & $[0,500]$ & No \\
6 & Fe\_Front\_3 & $[-105,105]$ & $[-105,105]$ & $[-105,105]$ & $[-105,105]$ & $[0,500]$ & No \\
\end{tabular}
\end{ruledtabular}
\end{table*}

\begin{table}[t]
\caption{\label{tab:odmrseeds}
Initial ODMR resonance frequencies used for pixel-wise Lorentzian fitting. Multiple values denote multiple initial resonance centers.
}
\begin{ruledtabular}
\begin{tabular}{llcc}
Defect & Dataset & Current-on seeds (GHz) & Current-off seeds (GHz) \\
\hline
1 & Cu\_Front\_2 & 2.755 & 2.762 \\
2 & Cu\_Front\_1 & 2.940, 2.942, 2.944 & 2.957, 2.959, 2.961 \\
3 & Cu\_Back\_12 & 2.927 & 2.921, 2.933 \\
4 & Fe\_Front\_1 & 2.938 & 2.935 \\
5 & Fe\_Front\_2 & 2.926 & 2.928 \\
6 & Fe\_Front\_3 & 2.955 & 2.955 \\
\end{tabular}
\end{ruledtabular}
\end{table}

\subsection{Uncertainty of fitted depth and dipole magnitude}
The covariance matrix of the profiled position fit was estimated from the Jacobian $J_p$ of the weighted residual with respect to $(x_0,y_0,z_0)$ at the optimum,
\begin{eqnarray}
\Sigma_p=\chi_\nu^2\left(J_p^T J_p\right)^{+}.
\end{eqnarray}
The $2\times2$ lateral covariance matrix $\Sigma_{xy}$ is the upper-left block of $\Sigma_p$. The 95\% lateral confidence ellipse is defined by
\begin{eqnarray}
\Delta\mathbf{r}_{xy}^{T}\Sigma_{xy}^{-1}\Delta\mathbf{r}_{xy}
=\chi^2_{2,0.95}=5.9914645,
\end{eqnarray}
where $\Delta\mathbf{r}_{xy}=(x-x_0,y-y_0)^T$. If
\begin{eqnarray}
\Sigma_{xy}=U\,\mathrm{diag}(\lambda_1,\lambda_2)U^T,
\end{eqnarray}
the sampled boundary points are
\begin{eqnarray}
\mathbf{r}^{(k)}_{xy}=\mathbf{r}_{0,xy}
+U
\begin{pmatrix}
\sqrt{\chi^2_{2,0.95}\lambda_1} & 0\\
0 & \sqrt{\chi^2_{2,0.95}\lambda_2}
\end{pmatrix}
\begin{pmatrix}
\cos\theta_k\\
\sin\theta_k
\end{pmatrix},
\end{eqnarray}
with $\theta_k=2\pi k/N_b$, $k=0,\ldots,N_b-1$, and $N_b=72$. The ellipse is not clipped to the fit ROI or to the nominal $x_0$ and $y_0$ bounds; this avoids artificially narrowing the propagated uncertainty when the confidence region extends outside those analysis windows.

At each boundary point $k$, $(x,y)$ is fixed at $\mathbf{r}^{(k)}_{xy}$, $z$ is reoptimized within the same sample-specific $z_0$ bounds used for the nominal fit, and $(Q_x,Q_y)$ are recalculated by weighted linear least squares,
\begin{eqnarray}
z^{(k)}&=&\underset{z_{\rm min}\le z\le z_{\rm max}}{\arg\min}\;\chi^2(x^{(k)},y^{(k)},z),\\
\mathbf{Q}^{(k)}&=&\widehat{\mathbf{Q}}(x^{(k)},y^{(k)},z^{(k)}).
\end{eqnarray}
The asymmetric fit intervals are defined by the extrema over the nominal solution and all 72 boundary refits,
\begin{eqnarray}
z_{\rm low}&=&\min\{z_0,z^{(1)},\ldots,z^{(N_b)}\},\\
z_{\rm high}&=&\max\{z_0,z^{(1)},\ldots,z^{(N_b)}\},\\
Q_{\rm low}&=&\min\{|Q_0|,|Q^{(1)}|,\ldots,|Q^{(N_b)}|\},\\
Q_{\rm high}&=&\max\{|Q_0|,|Q^{(1)}|,\ldots,|Q^{(N_b)}|\}.
\end{eqnarray}
Thus, $z_0$ is reported with errors $-(z_0-z_{\rm low})$ and $+(z_{\rm high}-z_0)$, and $|\vec{Q}|$ is treated analogously. The same ellipse boundary is drawn as the dashed confidence contour in the dipole-fit panels. For the copper back-side defect~\hyperlink{defect3}{3}, the confidence region reaches the depth bounds; the corresponding depth interval is therefore bound-limited.
\subsection{Effective-volume uncertainty and systematic effects}
The nominal effective volume is
\begin{eqnarray}
V_{\rm eff}=\frac{|Q_0|}{J_{\rm nom}},\qquad J_{\rm nom}=1.0\times10^6\ \mathrm{A/m^2}.
\end{eqnarray}
For the plotted fit uncertainty, $J_{\rm nom}$ is held fixed and only the ellipse-derived interval in $|\vec{Q}|$ is propagated,
\begin{eqnarray}
V_{\rm low}^{\rm fit}=\frac{Q_{\rm low}}{J_{\rm nom}},\qquad
V_{\rm high}^{\rm fit}=\frac{Q_{\rm high}}{J_{\rm nom}}.
\end{eqnarray}
The plotted intervals contain only fit-derived uncertainties. Uncertainty in the local current density and sensor--sample separation is treated as systematic uncertainty and is discussed in the main text.

The fitted parameters and fit-derived intervals are listed in Table~\ref{tab:fitresults}.

\begin{table*}[t]
\caption{\label{tab:fitresults}
Dipole-fit results. The intervals were obtained from the refits along the boundary of the 95\% lateral confidence ellipse and include only fit-derived uncertainty. $V_{\rm eff}$ was calculated using $J_{\rm nom}=1.0\times10^6\,\mathrm{A/m^2}$.
}
\begin{ruledtabular}
\begin{tabular}{lccccc}
Defect & Dataset & $z_0$ ($\mu$m) & Fit-derived interval ($\mu$m) & $V_{\rm eff}$ ($\mu$m$^3$) & Fit-derived interval ($\mu$m$^3$) \\
\hline
1 & Cu\_Front\_2 & 513 & $[499,530]$ & $1.57\times10^8$ & $[1.45,1.72]\times10^8$ \\
2 & Cu\_Front\_1 & 154 & $[145,173]$ & $1.97\times10^6$ & $[1.72,2.68]\times10^6$ \\
3 & Cu\_Back\_12 & 439 & $[400,2000]$ & $7.49\times10^7$ & $[5.39\times10^7,6.51\times10^9]$ \\
4 & Fe\_Front\_1 & 83.6 & $[80.4,86.9]$ & $4.69\times10^6$ & $[4.34,5.07]\times10^6$ \\
5 & Fe\_Front\_2 & 98.8 & $[97.6,100]$ & $3.81\times10^6$ & $[3.72,3.90]\times10^6$ \\
6 & Fe\_Front\_3 & 101 & $[99.4,102]$ & $5.47\times10^6$ & $[5.33,5.61]\times10^6$ \\
\end{tabular}
\end{ruledtabular}
\end{table*}

\end{document}